\documentclass[journal]{IEEEtran}
\IEEEoverridecommandlockouts
\usepackage{amsmath,amssymb,amsfonts}
\usepackage{algorithmic}
\usepackage{graphicx}
\usepackage{textcomp}
\usepackage{xcolor}
\usepackage[utf8]{inputenc}
\usepackage[style=apa,backend=biber]{biblatex} 
\def\BibTeX{{\rm B\kern-.05em{\sc i\kern-.025em b}\kern-.08em
    T\kern-.1667em\lower.7ex\hbox{E}\kern-.125emX}}
    \makeatletter

\begin{document}

\title{Biomechanical Analysis of Breast Cancer Cells: \\A Comparative Study of Invasive and Non-Invasive Cell Lines Using Digital Holographic Microscopy\\}
 \author{
    \IEEEauthorblockN{1\textsuperscript{st} Hasan Berkay ABDIOGLU\(^1\)}, \IEEEauthorblockN{2\textsuperscript{nd} Yagmur ISIK\(^1\)}, \IEEEauthorblockN{3\textsuperscript{rd} Merve SEVGI\(^2\)}, \IEEEauthorblockN{4\textsuperscript{th} Ufuk Gorkem KIRABALI\(^1\)}, \\ \IEEEauthorblockN{5\textsuperscript{th} Caner KARACA\(^3\)}, \IEEEauthorblockN{6\textsuperscript{th} Yunus Emre MERT\(^1\)}, \IEEEauthorblockN{7\textsuperscript{th} Defne DEDEHAYIR\(^4\)}, \IEEEauthorblockN{8\textsuperscript{th} Berke Tuna BOSTANCI\(^5\)}, 
    \\ \IEEEauthorblockN{9\textsuperscript{th} Yasemin BASBINAR\(^3\)}, \IEEEauthorblockN{10\textsuperscript{th} Huseyin UVET\(^1\)\\}
     
    \IEEEauthorblockA{\textit{\\\(^1\)Department of Mechatronics Engineering, Yildiz Technical University, Istanbul, Turkey\\}}
    \IEEEauthorblockA{\textit{\(^2\)Department of Bioengineering, Yildiz Technical University, Istanbul, Turkey\\}}
    \IEEEauthorblockA{\textit{\(^3\)Department of Translational Oncology, Dokuz Eylul University, Izmir, Turkey\\}}
    \IEEEauthorblockA{\textit{\(^4\)Hisar School, Istanbul, Turkey}}
    \IEEEauthorblockA{\textit{\(^5\)The Sezin School, Istanbul, Turkey}}
   
}

\maketitle

\begin{abstract}
This study investigates the biomechanical properties of breast cancer cells, focusing on the invasive MDA-MB-231 and non-invasive MCF-7 cell lines, using phase-shifting digital holographic microscopy and an acousto-holographic system for non-invasive strain measurement. Findings reveal higher strain values in MDA-MB-231 cells, indicative of their aggressive, metastatic nature, compared to lower strain values in MCF-7 cells. This variance in mechanical properties, potentially linked to distinct metabolic states and responses to external stimuli, underscores the role of cellular mechanics in cancer progression. The study advances understanding of breast cancer cell mechanics, highlighting biomechanical analysis as a crucial tool in cancer research and potential therapeutic interventions.\\
\end{abstract}


\section{Introduction}
Breast cancer is a malignancy characterized by the abnormal transformation and uncontrolled proliferation of cells constituting breast tissue \parencite{who2023}. It ranks among the most prevalent cancer types globally. Breast cancer is prevalent among women, though it can also occur, albeit rarely, in men \parencite{who2023}. The higher incidence of breast cancer in women compared to men can be attributed to several factors, including the continuous exposure of breast tissue to hormones like estrogen and progesterone, which predispose it to growth \parencite{lukasiewicz2021}. The frequency of breast cancer occurrence in an individual is also influenced by hormonal factors, genetic predisposition, and lifestyle choices \parencite{lukasiewicz2021}.
Breast cancer is categorized into three main types: noninvasive (in situ), invasive and metastatic breast cancers. Invasive breast cancer has the potential to metastasize, whereas noninvasive cancer does not exhibit spreading characteristics \parencite{feng2018}. Within the noninvasive category, there are two subtypes: ductal carcinoma in situ and lobular carcinoma in situ. Lobular carcinoma in situ involves an increase in lobular cells within the breast tissue and can elevate the risk of breast cancer \parencite{makki2015}. In such cases, close monitoring and prophylactic treatments, including surgical interventions, may be considered. The diagnosis and treatment of breast cancer are tailored to the patient's individual factors. Early detection plays a crucial role in the effective treatment of this cancer type \parencite{ginsburg2020}. Ductal carcinoma in situ typically does not present symptoms during examination and possesses the potential to transition from normal to cancerous cells. It is commonly marked with wires and radioactive substances during surgical procedures for removal \parencite{tomlinsonhansen2023}.
Energy metabolism in cancer cells adapts during progression \parencite{hanahan2011, vaupel2021}. These metabolic alterations encourage cellular behavior from mitosis to metastasis. The infamous Warburg effect (aerobic glycolysis) defines that proliferative cells often run glycolysis and its branched pathways to support progression, even when oxygen is sufficient \parencite{vaupel2021}. However, not all cancer cells reorient their metabolism through the Warburg effect. Several mutations and epigenetic modifications alter oxidative phosphorylation to take advantage of the Pasteur phenotype \parencite{karaca2022}. Both Warburg and Pasteur energitic subtypes could robust some metastatic pathways and phenotypes, but in different ways \parencite{lu2019, kamarajugadda2012,karaca2022}.
Triple-negative MDA-MB-231 and estrogen-responsive MCF-7 cell lines adapt their energy metabolism differently. While MCF-7 is a Pasteur subtype, the more aggressive MDA-MB-231 has a Warburg phenotype. Recent studies indicate that their metabolic state also influences their metastatic behavior. Normalization of energy metabolism by site-directed expression of pyruvate dehydrogenase enzyme provide MDA-MB-231 susceptible to anoikis and inhibits its metastatic behavior \parencite{kamarajugadda2012}. The energetic subtype may also influence homing. A recent study suggests that calcification of the OXPHOS dominant cell line MCF-7 induces TGF-\(\beta\) dependent bone metastasis \parencite{hu2020}. These different metastatic routes could be related to various cell stiffness criteria due to cell cytoskeleton reorganization, ROS generation, and membrane lipid composition \parencite{dewane2021,sun2014,szlasa2020}.
The cytoskeleton, comprising actin filaments, microtubules, and intermediate filaments, significantly influences cellular mechanics \parencite{mandal2016}. Alterations in these cytoskeletal fibers directly impact cell mechanics. Cellular mechanics, which are crucial in understanding various cellular functions such as cell migration, adhesion, and division, undergo significant changes during severe diseases like cancer \parencite{laverde2021,mandal2016}. One key mechanical property, cellular strain, is closely linked to cell motility \parencite{lu2019}. Cancer cells exhibit higher strain values compared to normal cells, aligning with their need for movement between tissues \parencite{zoellner2015}. Strain values vary among cancer cells with different levels of aggressiveness. For instance, MCF-7 cells, which have a lower invasive capability, show different strain values compared to the highly invasive MDA-MB-231 cancer cells, with MDA-MB-231 exhibiting higher strain \parencite{zbiral2023}.
Cell strain, a fundamental parameter in understanding cellular biomechanics, has been measured using various techniques for a long time. Although Atomic Force Microscopy (AFM) is considered the gold standard, its limitations include low resolution, time-consuming measurements, and an invasive nature. In recent years, digital holographic microscopy has emerged as a prevalent method in cell imaging. Varol et al. developed an acousto-holographic system, demonstrating that cell strain measurements can be conducted in a non-invasive and label-free manner.
In this study, the acoustic wave frequency used by Varol et al. was reduced, and a higher fps (frames per second) was achieved, thereby increasing the sample size. This adjustment allows for more detailed and accurate measurements of cell strain, enhancing the understanding of cellular biomechanics, particularly in the context of cancer cell behavior and properties.

\section{MATERIALS AND METHODS}

\subsection{Materials}

For the invasive breast cancer cell line MDA-MB-231 (ATCC, HTB-22) and the non-invasive breast cancer cell line MCF-7, DMEM-F12 (Gibco, 21331020) supplemented with 10\% fetal bovine serum and 1\% Penicillin-Streptomycin will be used. The Sylgard 184 Silicone Elastomer Kit (Dow) will be employed for the production of PDMS chips.

\subsection{Cell Culture}
Cells will be passaged when they reach 70-80\% confluency. For passaging, cells will be washed with 2 mL of sterile Phosphate Buffered Saline (PBS) (pH 7.4) and then treated with 500 \(\mu\)l of 0.25\% Trypsin/EDTA solution. The flask will be incubated for 5 minutes at 37°C in a 5\% CO2 incubator (Thermo Scientific, Germany). Detached cells will be transferred to a falcon tube and centrifuged at 1000 rpm for 5 minutes. After centrifugation, the supernatant will be discarded, and 1 \(\mu\)l of the cell pellet will be mixed with 49 \(\mu\)l of trypan blue and added to a Thoma slide for counting under a microscope. Cell seeding will be performed in each flask to achieve a density of \(1.5×10^5\) cells. Seeded cells will be maintained in a 37°C incubator with 5\% CO2 \parencite{erci2018green}.

\subsection{Production of Transducer-Integrated PDMS Chip}
The design of the microfluidic chip and channels will be done using 3D design software. The microfluidic device will be fabricated using Poly(dimethylsiloxane) (PDMS) with dimensions of 40 mm × 24 mm × 5 mm using UV lithography. The process includes cleaning of silicon wafers, spinning of SU-8 coating, exposure to UV, and development in Propylene glycol methyl ether acetate (PGMEA). The microfluidic chip will have a channel designed as 10 mm x 10 mm square for cell seeding. For chip production, silicone elastomer and curing agent will be mixed in a specified ratio and poured into a mold. To remove bubbles, the PDMS will be left in a desiccator and then cured at 80°C for 2 hours. The cured PDMS will be cut from the main mold using a scalpel. Finally, the PDMS will be exposed to oxygen plasma at 4 mbar for 120 seconds for bonding with glass.

\subsection{Cell Seeding on Chips}
MCF-7 and MDA-MB-231 cells, counted using trypan blue on a Thoma slide after trypsinization, will be seeded onto the transducer-integrated PDMS chips. The seeded cells will be incubated in their respective media for 48 hours in a 37°C incubator with 5\% CO2.

\subsection{Experimental Setup}
In this study, we employed a highly specialized experimental setup centered around a Mach-Zehnder interferometer with phase-shifting capabilities to facilitate holographic imaging.  We used a consistent light source, a 671 nm, 10 mW DPSSL laser. The system divided incident light into two beams: a reference beam and an object beam, with phase modulation achieved in the reference beam through a precision PZT actuator (New Focus, Picomotor 8302). A microfluidic chamber containing cell culture (or reference microbeads) is placed in the other arm of the interferometer, and a PZT transducer (SMBA25W7T05PV, Steiner Martins Inc.) provides acoustic stimulation at approximately 10 Hz frequency. The light fields are captured by microscope objective lenses (Newport M-20X, 20x magnification) and recombined at the second beam splitter of the interferometer after passing through the phase-shifting device and the microfluidic chamber figure 1.  To ensure the system's accuracy, meticulous adjustments of mirror angles and beam splitter positions were made, ensuring optimal interference pattern visibility. Basic calibration involved the introduction of a matte surface in front of a CMOS camera to visualize interference patterns, creating distinct brightness differences between object and reference beams. Further calibration precision was achieved by aligning the interference pattern's center onto the camera. Micron-level focusing of cells was achieved using the Newport HXP50 Hexapod system. For image processing, interferograms were processed using the least squares method, allowing for robust handling of potential intensity variations resulting from diverse measurement methods and conditions. The core of our research involved the generation of controlled acoustic vibrations applied to cell surfaces, causing membrane vibrations proportional to the applied force. Phase-shifting digital holographic microscopy was employed for precise quantification of cell membrane displacement and elasticity, facilitated by continuous interferogram acquisition. To subject cells to controlled vibrations, a specialized acoustic transducer operating at 10 Hz with thickness modulation was utilized. This systematic process generated a comprehensive set of interferograms with varying phases, enabling a detailed characterization of cell responses during vibration \parencite{varol2022}.

\subsection{Processing of Images}
Interferograms are fundamentally photographic images that incorporate optical interference. Each pixel in an interferogram represents the intensity of the light beam. This intensity can be mathematically expressed by the electromagnetic wave equation as \( I'+I''cos(\phi(x,y)+\delta(t))\).
This formula encompasses three distinct unknown variables. To solve for these unknowns, a minimum of three equations is required. However, due to potential discrepancies in the measured intensities arising from the measurement method, the instrumentation used, and the specific location of measurement, processing more than three interferograms using the numerical method of least squares can yield results that are closer to the actual values. Variations in the phase change intervals will also lead to modifications in the equation that determines the phase value.
The criteria for the least squares method can be articulated as \(\frac{\delta S_j}{\delta a_j} = 0, \quad \frac{\delta S_j}{\delta b_j} = 0, \quad \frac{\delta S_j}{\delta c_j} = 0.\)
 This can be further simplified to \(X_j=A^{-1}/B_j\), where:

\begin{equation}
\mathbf{X}_j = \begin{pmatrix} {a_j , b_j , c_j} \end{pmatrix}^T\label{eq}
\end{equation}

\begin{equation}
\left[
\begin{array}{lll}
M & \sum\limits_{i=1}^{M} \cos(\delta_i) & \sum\limits_{i=1}^{M} \sin(\delta_i) \\
\sum\limits_{i=1}^{M} \cos(\delta_i) & \sum\limits_{i=1}^{M} \cos^2(\delta_i) & \sum\limits_{i=1}^{M} \cos(\delta_i)\sin(\delta_i) \\
\sum\limits_{i=1}^{M} \sin(\delta_i) & \sum\limits_{i=1}^{M} \sin(\delta_i)\cos(\delta_i) & \sum\limits_{i=1}^{M} \sin^2(\delta_i)
\end{array}
\right]
\end{equation}

\begin{equation}
\mathbf{B}_j = \begin{pmatrix}
\sum_{i=1}^{M} I_{ij} \;  \; \sum_{i=1}^{M} I_{ij} \cos(\delta_i) \;  \; \sum_{i=1}^{M} I_{ij}
\end{pmatrix}\label{eq}
\end{equation}

\begin{figure}[!]
    \centering
    \includegraphics[width=1\linewidth]{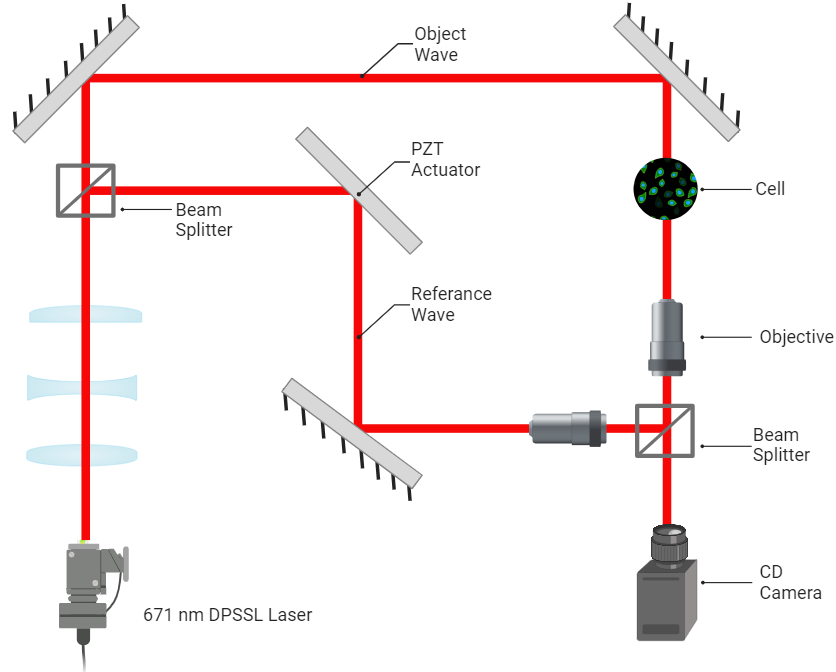}
    \caption{\textit{Schematic of the Mach-Zehnder Interferometry Setup. The beam from a light source is split into two by a beam splitter, with one beam designated as the reference beam and the other as the object beam. The phase of the reference beam is shifted using a PZT (lead zirconate titanate) actuator, creating a phase shift in the interference pattern. An objective lens is used to magnify both the object and reference beams equally. After the beams interfere with each other, the resulting pattern is captured by a CMOS (Complementary Metal-Oxide-Semiconductor) camera.} }
    \label{fig:mach_zehnder}
\end{figure}

\(\\\frac{\delta S_j}{\delta a_j} = 0, \quad \frac{\delta S_j}{\delta b_j} = 0, \quad \frac{\delta S_j}{\delta c_j} = 0.\)
represent the summation conditions for the least squares criteria.
" \(X_j=A^{-1}/B_j\)" denotes the solution for the unknowns in the system, with \(A^{-1}\) representing the inverse of matrix A, and \(Bj\) being a vector in the system of equations.
Using these equations, the values of \(a_{j }\) \(b_{j }\) and \(c_{j }\) can be calculated. Consequently, the phase value for each pixel can be given by \(\varphi_j=arctan((-c_j)/b_j )\) . Since the obtained phase values are based on the arctangent function, divergence is expected at certain values (multiples of \(pi/2\)). This divergence can disrupt the continuity of the phase map. To ensure continuity, phase unwrapping (Phase Unwrapping) should be performed. Ultimately, the optical path difference (OPD) can be calculated using the phase map obtained, with the help of the following formula \parencite{aninditha2019}:

\begin{equation}
OPD = \frac{\lambda}{2\pi} \\ \varphi_j\label{eq}
\end{equation}

Here, \(\lambda\) represents the wavelength of the light used in the interferogram, and \(\phi_j\) is the unwrapped phase value at each pixel. The optical path difference (OPD) provides crucial information about the optical properties of the material or the system under investigation.

\subsection{Reconstruction of Strain Maps }
In the setup designed for hardness measurement, a wave generator (Siglent SDG6032X) is used to produce waves. The generated signals are then amplified using a voltage amplifier (Falco Systems, WMA-300) before being sent to a PZT (lead zirconate titanate) transducer (Steminc SMBA25W73T05PV) for conversion into acoustic waves.
The generated acoustic waves induce acoustic pressure on the surface of cells, causing the cell membrane to oscillate in proportion to the acoustic force. The resulting cell membrane oscillations are observed using a high-speed camera (CMOS Camera). From these observations, the displacement of the cell membrane is measured. The strain value of the cell is determined by calculating the ratio of displacement to the original depth.
This method leverages the principles of acoustics and mechanics to non-invasively probe the mechanical properties of cells. By analyzing the cell's response to the induced acoustic waves, valuable insights into its mechanical properties, such as strain, can be obtained. This information is crucial in various fields, including biomedical research and material science, where understanding the mechanical properties of cells is essential.
In this setup, the vibration of the cell membrane is measured using phase-shift digital holographic microscopy with a Mach-Zehnder arrangement. This method aims to continuously obtain interferograms. The different phases of the obtained interferograms enable the 3D modeling of the cell.
To induce vibrations in the cells, a 10 Hz thickness-mode acoustic transducer is used. The vibration motion of the transducer generates acoustic waves on PDMS chips, causing the cells to vibrate. It is assumed that the cell vibrates with a periodic stimulus at a frequency of \(f_{0}\). To capture interferograms at desired points during the vibration period, the camera is triggered using a periodic frequency signal of  \(f_{0}/(1 + f_{0}/\delta t)\). After \(\delta t / f_{0}\)
 cycles, a series of interferograms is obtained for time shifts of \(k \delta t\) for \(k = 0, 1, \ldots, n\) (\(n \delta t = \frac{1}{f_{0}}\)). This process is repeated for many time intervals, each yielding a series of interferograms at different phases \parencite{varol2022}.
To determine cell strain, a mechanical model that can describe the mechanical deformation induced on the cell membrane by the vibration of the piezoelectric transducer is assumed. Strain is defined as relative deformation, compared to a reference position configuration. For modeling the small-scale deformations in the membranes of adherent cells, this formula can be adapted to measure the displacement of the cell membrane due to acoustic stimulilation. Cell membrane is subjected to a force that causes it to deform, the strain would be the ratio of the displacement of the membrane (how much it stretches or compresses) to its original, unstressed length.

\begin{equation}
\epsilon = \frac{\Delta L}{L_0}\label{eq}
\end{equation}

In cell biomechanics, strain \(\epsilon\) is calculated as \( = \frac{\Delta L}{L_0}\)
  , where \(\Delta L\) is the cell membrane's displacement and \(L_0\) is its original length. This formula measures how much the membrane stretches or compresses in response to external forces, providing insights into cellular responses to acoustic stimuli.

  \begin{figure}[t]
    \centering
    \includegraphics[width=1\linewidth]{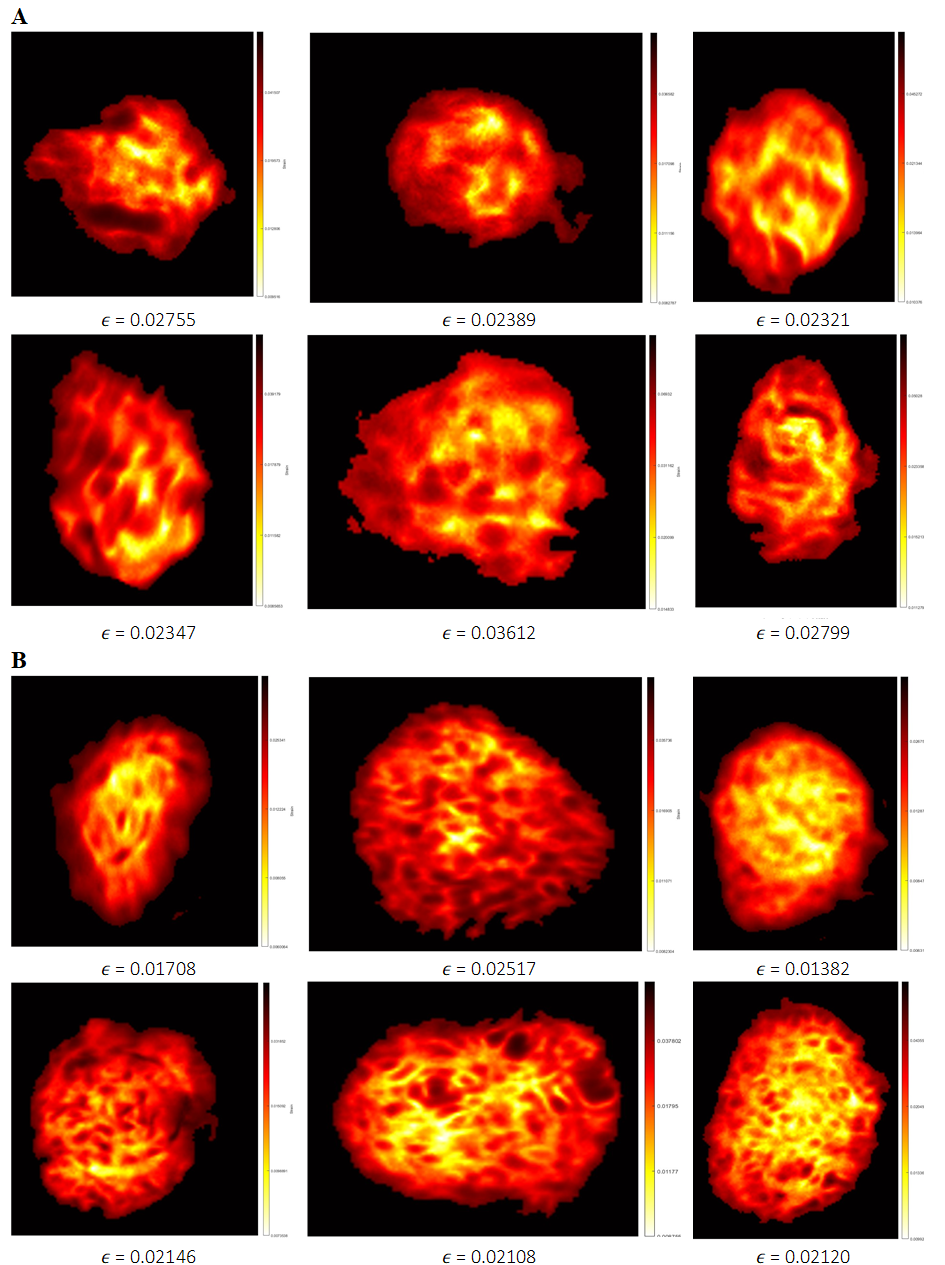}
    \caption{\textit{This figure illustrates the distribution of strain values \(\epsilon\) in two distinct breast cancer cell lines: MDA-MB-231 (Panel A) and MCF-7 (Panel B). Each cell within the panels represents a specific strain measurement from individual experiments. The MDA-MB-231 cells, show a range of strain values, with some cells exhibiting notably higher strain, indicative of their metastatic potential. In contrast, the MCF-7 cells, display generally lower strain values.\\}}
    \label{fig:enter-label}
\end{figure}

\section{result}

In this study, we conducted a total of 14 separate experiments for each of the breast cancer cell lines, MDA-MB-231 and MCF-7. The outcomes of these experiments are detailed in Table 1. figure 2-a presents the strain measurements of MDA-MB-231 cells, captured through our holographic imaging setup. These measurements represent the strain values for fourteen distinct cells over the observation period. In a similar vein, the strain measurements for MCF-7 cells are depicted in figure 2-b.

\begin{table}[!]
\caption{\textit {Comparative Strain Measurements of Breast Cancer Cell Lines MDA-MB-231 and MCF-7. In this table displays the strain values for each of the 14 experiments conducted on the MDA-MB-231 and MCF-7 breast cancer cell lines. The values are indicative of the biomechanical properties of these cells, with MDA-MB-231 representing a more aggressive phenotype and MCF-7 a less invasive one.}}
  \centering
  \renewcommand{\arraystretch}{1.5} 
  \setlength{\tabcolsep}{10pt} 
  \begin{tabular}{ccccccc}
    \textbf{No} & \textbf{MDA} & \textbf{MCF} & \textbf{No} & \textbf{MDA} & \textbf{MCF} \\
    \hline
    1 & 0.02321 & 0.01368 & 8 & 0.02755 & 0.01833 \\
    2 & 0.02347 & 0.01382 & 9 & 0.02799 & 0.02108 \\
    3 & 0.02364 & 0.01453 & 10 & 0.03095 & 0.0212 \\
    4 & 0.02389 & 0.01524 & 11 & 0.03348 & 0.02146 \\
    5 & 0.02461 & 0.01708 & 12 & 0.03398 & 0.02176 \\
    6 & 0.02522 & 0.01732 & 13 & 0.03408 & 0.02517 \\
    7 & 0.02568 & 0.01798 & 14 & 0.03612 & 0.02872 \\
    \hline

  \\\end{tabular}
  
  \label{tab1}
\end{table}

Furthermore, the calculated parameters of the strain model are illustrated in figure 3. An intriguing observation from these results is the apparent correlation between strain values and the potential for metastasis. The data indicates that the mean strain value for the MDA-MB-231 cells, which are known for their more aggressive and metastatic nature, was found to be 0.02813357. In contrast, the mean strain for the less invasive MCF-7 cells was lower, recorded at 0.01909786. This variance in strain values between the two cell lines underscores the potential link between cellular biomechanical properties and their metastatic capabilities.

\begin{figure}[!]
    \centering
    \includegraphics[width=1\linewidth]{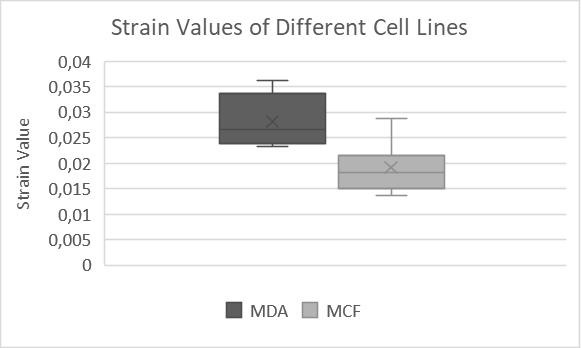}
    \caption{\textit{This figure presents a comparative analysis of the strain values for the MDA-MB-231 and MCF-7 breast cancer cell lines. It highlights the average biomechanical strain exhibited by each cell line, illustrating the distinct mechanical properties associated with their differing levels of invasiveness and metastatic potential.}}
    \label{fig:enter-label}
\end{figure}

\section{DISCUSSION}
This study has provided significant insights into the biomechanical properties of breast cancer cells, specifically focusing on the invasive MDA-MB-231 and the non-invasive MCF-7 cell lines. Through the innovative use of phase-shifting digital holographic microscopy and a specialized acousto-holographic system, we have been able to non-invasively measure and compare the strain values of these two cell lines under acoustic stimulation.
Our findings reveal that the MDA-MB-231 cells, characterized by a Warburg phenotype, exhibit higher strain values compared to the MCF-7 cells, which align with a Pasteur subtype. Specifically, the strain values for MDA-MB-231 ranged from 0.02321 to 0.03612, while those for MCF-7 were between 0.01368 and 0.02872. This difference in strain values is indicative of the distinct mechanical properties and possibly the metastatic potential of these cell lines. The higher strain values in MDA-MB-231 cells correlate with their more aggressive nature and higher invasive capabilities.
The study underscores the importance of cellular biomechanics in cancer research. The differences in strain values between the two cell lines could be linked to their distinct metabolic states and their response to external stimuli, such as acoustic waves. This suggests that cellular mechanics, alongside genetic and molecular profiles, play a crucial role in the progression and behavior of cancer cells.
In conclusion, our study not only advances the understanding of the mechanical properties of breast cancer cells but also highlights the potential of biomechanical analysis as a tool in cancer research and therapy. The differences in strain values between MDA-MB-231 and MCF-7 cells provide valuable insights into their respective mechanical properties and behaviors, which could be pivotal in developing targeted treatments and interventions for breast cancer.\\\\

\printbibliography 
\end{document}